# Noise performance of the complex monopulse ratio

Marco Lanucara

***Abstract*—The paper provides a characterization of the complex monopulse ratio in terms of autocorrelation and power spectral density of its fluctuations during satellite tracking, taking into account the presence of additive noise on sum and difference channels. The considered spectral structure and statistical distribution of the incoming signal is of interest for satellite missions. In particular it is assumed that the signal available at the monopulse processor after frequency down conversion contains a Gaussian term produced by low pass filtering of a constant envelope modulation, plus a monochromatic component representative of a possible residual carrier. The results can be used for optimizing the design of a monopulse tracking system.**

## I. Introduction

SATELLITE tracking by ground directive antennas requires accurate pointing to the flying system. For example, the half-power beamwidth of an antenna with a diameter of 35m operating at 32GHz can be as small as 17 to 19 millidegrees, which means that a pointing accuracy in the order of few millidegrees will be required, in order not to penalise the communications link budget. The pointing to the satellite can be achieved either by programming the antenna direction based on the known position of the satellite (blind pointing), or by use of the received signal for tracking purposes. The latter method, which is the subject of this paper, is especially useful when the a-priori position of the satellite is not known with sufficient accuracy, e.g. shortly after launch, at the first acquisition of signal, or during "routine" tracking when the blind pointing leads to a residual slowly varying pointing error. There are various techniques by which a received signal can be used for tracking purposes, however in this context we are interested in monopulse autotrack [1], where the antenna feed system produces error signals based on the signal reception, without conical scanning. In particular a monopulse autotrack system produces two signals, called "sum" and "difference" (or "delta"), which encode the angular misalignment between the antenna boresight and the direction of the satellite: more specifically and with reference to Fig. 1, if $\theta_s$ and $\varphi_s$ are the two angles identifying the satellite direction in the antenna frame, and if $\theta_s$ is a fraction of the half-power beamwidth, two complex signals $\Sigma(t)$ and $\Delta(t)$ will be available at the input of the monopulse processing system after frequency down conversion to complex baseband and low pass filtering, related by the following

$$\Delta(t) = \theta_s K_F e^{i\varphi_s} \Sigma(t). \tag{1}$$

The signal $\Sigma(t)$ is used not only for tracking, its unfiltered version carries the information transmitted by the satellite, suitably coded and modulated onto the downlink carrier. The signal $\Delta(t)$ carries the tracking information, it vanishes when the tracking is perfect and its amplitude and phase, referred to $\Sigma(t)$, contain the components of the angular offset. The constant $K_F$ is a parameter of the antenna system (the so called "tracking slope"), and determines the magnitude of the difference signal for a given angular misalignment. It is to be remarked that the Eq. (1) is ideal in many respects: first of all noise is present in a real system, added to both signals $\Delta(t)$ and $\Sigma(t)$; secondly different time delays can be experienced by the two signals during the frequency down-conversion process, such that a spurious phase term can show up on top of $\varphi_s$; furthermore the error signal may not vanish even while perfectly tracking, and the tracking null may be displaced vs. the antenna boresight direction. All above effects (plus other) result in a degradation of the tracking performance, and will be ignored in this paper, with the exception of the first, the presence of noise, which can never be eliminated. When considering noise, the signals available at the input of the monopulse processing system can be expressed as follows

$$\begin{aligned} S(t) &= \Sigma(t) + n_\Sigma(t), \\ D(t) &= \theta_s K_F e^{i\varphi_s} \Sigma(t) + n_\Delta(t) \end{aligned} \tag{2}$$

where $n_\Sigma(t)$ and $n_\Delta(t)$ are complex noise terms. Let's now consider the simple action performed by the monopulse processor: it forms the ratio between the noisy difference and sum signals

$$M(t) = \frac{D(t)}{S(t)} \tag{3}$$

and integrates it over a configurable time interval $T$, e.g.

$$M_T(t) = \frac{1}{T} \int_{t-T}^{t} M(t')\, dt'. \tag{4}$$

Marco Lanucara is with the European Space Agency.

The details of the time integration may differ from Eq. (4), however they are unimportant in this context, the main point being that the time integration will act as a low pass filter on $M$. Indeed, once the spectral properties of $M$ are known, the ones of $M_T$ may be derived by knowledge of the specific filter transfer function. For such reason, in the following we will focus on $M$.

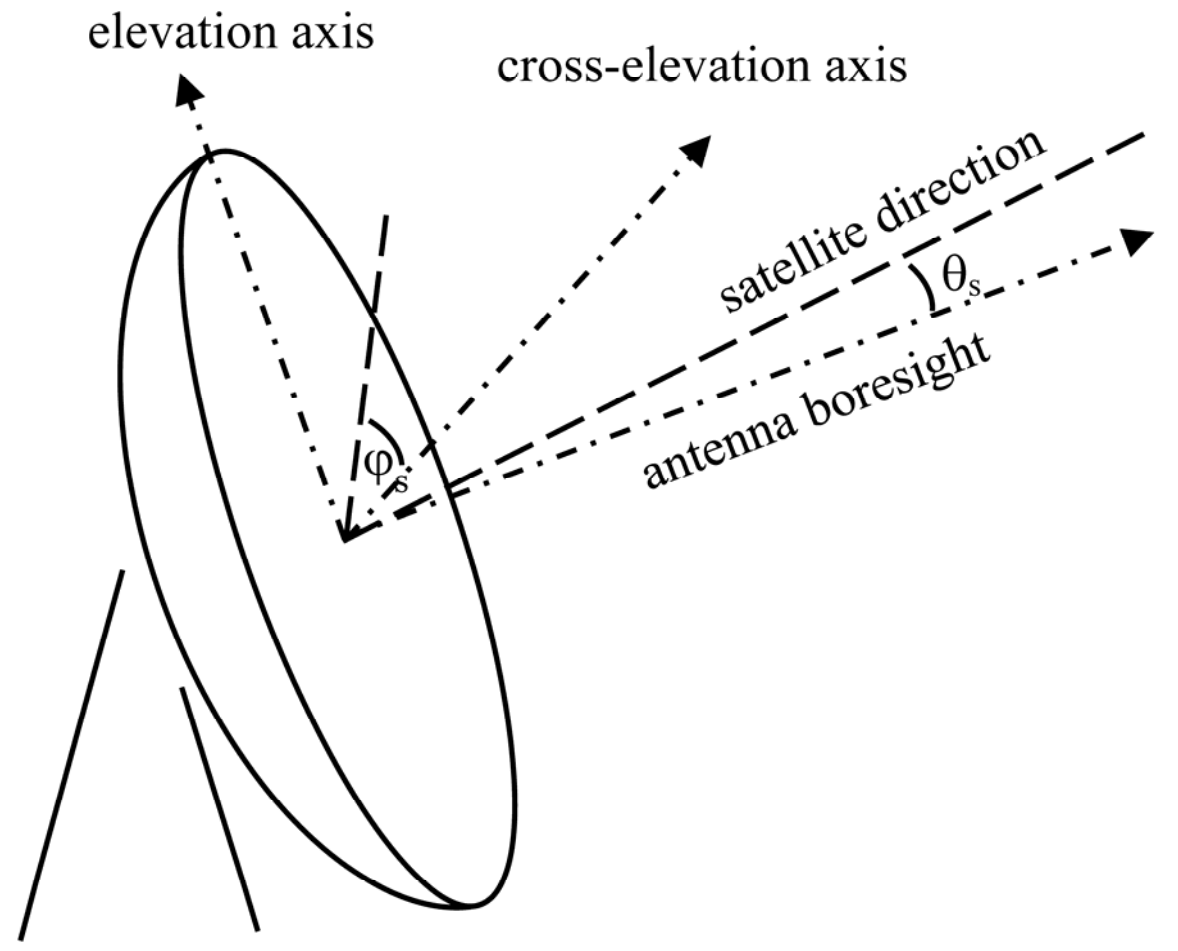

Fig. 1: definition of antenna coordinate frames and representation of the angular misalignment between antenna boresight and satellite direction through the angles $\theta_s$ and $\varphi_s$.

First of all we notice that, in absence of noise, Eqs. (2) and (3) give

$$(M)_{\text{no noise}} = \theta_s K_F e^{i\varphi_s}, \quad (5)$$

which shows that, after dividing by the known $K_F$, the complex ratio contains the complete information about the angular misalignment between antenna boresight and satellite direction, and the error signal, split over its real and imaginary components, could be fed back to a servo system to enable autotrack. In presence of noise however the magnitude of the mean of the complex ratio will decrease, and $M$ will fluctuate around its complex mean. In general, under noisy conditions, one may write

$$M = \alpha \cdot (M)_{\text{no noise}} + \delta M \quad (6)$$

where $0 \leq \alpha \leq 1$ is a reduction factor and $\delta M$ is the fluctuation of the ratio around the reduced mean. The expression of $\alpha$ can be retrieved from literature [2], together with the full expression of the probability density function of the ratio. The subject of this paper is finding the spectral density of $\delta M$, whose knowledge is mandatory for designing an autotrack feedback loop, for a class of signals spectra and statistical distributions of interest for applications. The expression of such parameter is known for purely Gaussian signal and noise (see for example [3] for what concerns the correlation properties of $\delta M$), however the addition of a residual carrier leads to a more general expression, which is analysed in this paper. Before starting the analysis, some simplifying assumptions are needed. First of all we have to consider that the time integration in Eq. (4) will select only the low frequency components of $M$, the reason being that its mean value, which is the quantity of interest, is slowly varying because it is linked to a physical misalignment between two directions. For this reason the knowledge of the spectral density of $\delta M$ is required only around the zero frequency of the complex baseband. The second consideration is that during tracking, which is the regime of interest, the real and imaginary components of the monopulse ratio fluctuate around a zero mean, which will substantially simplify the computation of the spectral density of $\delta M$.

## II. SIGNALS MODEL

Due to the variety of possible carrier modulations, it is not possible to establish a priori what kind of spectral structure and statistical distribution could be applied to $\Sigma(t)$. A constant envelope modulation is desirable in many satellite applications, due to the need of operating the on-board power amplifier near saturation, where its power efficiency is maximum [4]. However low pass filtering is generally applied prior to monopulse processing, to increase the signal to noise ratio at the output of the filter. Due to the filtering, the constant envelope condition will generally be lost, and one could model the signal $\Sigma(t)$ as a complex Gaussian process. Such modelling however would be insufficient when dealing with modulations with residual carrier, still common especially in deep space applications. Instead an adequate modelling of $\Sigma(t)$ could be the following

$$\Sigma(t) = \sqrt{P_c} e^{i(2\pi f_c t + \theta_c)} + x(t), \quad (7)$$

where $P_c$, $f_c$ and $\theta_c$ are the power, frequency and initial phase of the residual carrier and $x(t)$ is the realization of a complex zero mean Gaussian process of power $P_x$ representative of the frequency down-converted and low-pass filtered modulation spectrum. When including the noise on the sum channel as per Eq. (2), we have the following expression for $S(t)$

$$S(t) = \sqrt{P_c} e^{i(2\pi f_c t + \theta_c)} + x(t) + n_\Sigma(t), \quad (8)$$

i.e. the sum of a monochromatic signal plus a Gaussian process including noise, of power $P_\Sigma$, and signal component. In summary, the spectrum of $S(t)$, available to the monopulse processing system may look like the example of Fig. 2.

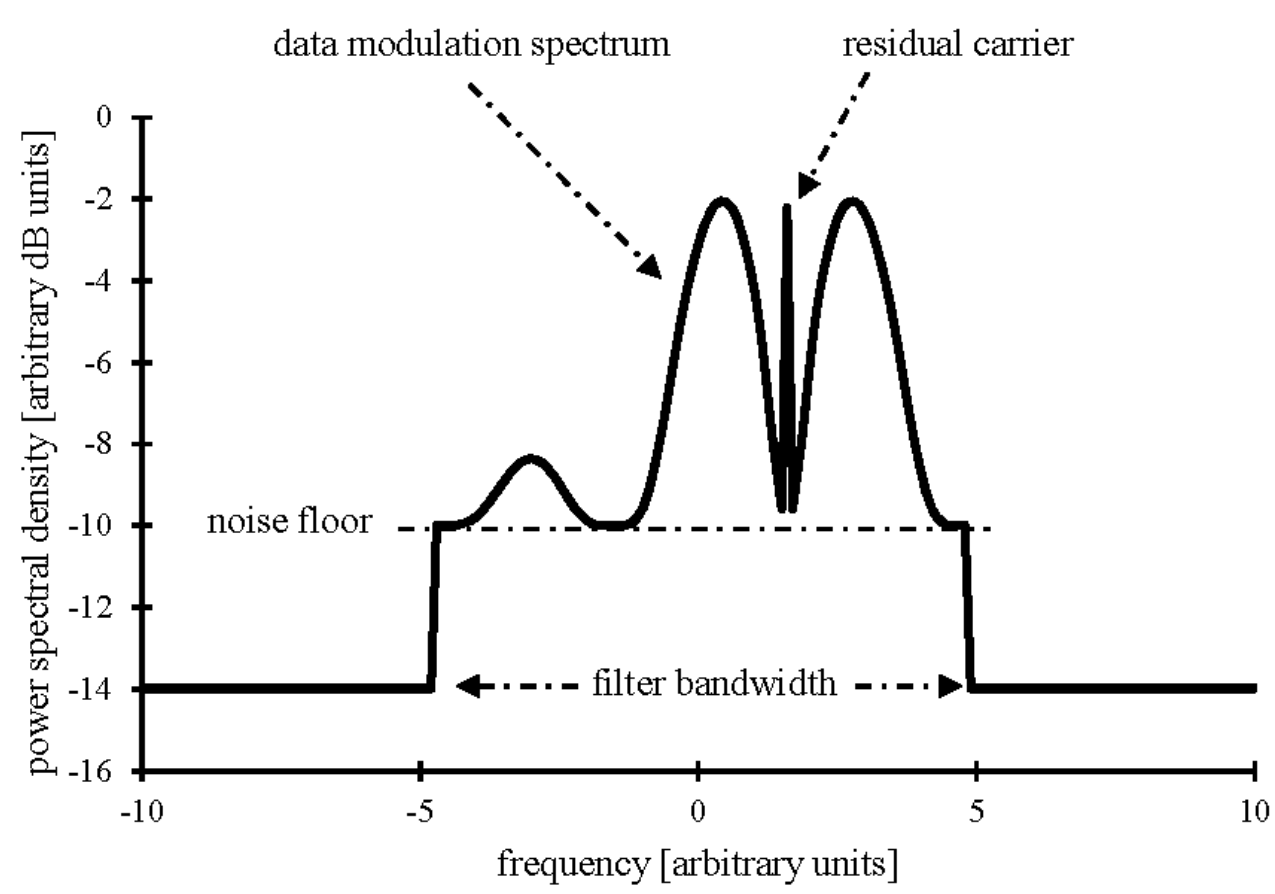

Fig. 2: sample power spectrum at the sum input of the monopulse processor

Concerning the complex noise terms $n_\Delta(t)$ and $n_\Sigma(t)$, we assume them to be realizations of zero-mean Gaussian processes statistically independent between each other and with $x(t)$. It is to be remarked that in other applications like radar, the statistical independence between sum and difference noise may be inadequate, as in those contexts interfering signals are embedded in the noise terms. However in the satellite tracking scenarios under consideration the correlation between sum and difference noise reduces to a weak component linked to the sky brightness, which can generally be neglected.

To complete the modelling, we report some properties of the autocorrelation of the involved signals, which we denote generically with $\xi(t)$ until the end of the section. We adopt in this paper the following definition of autocorrelation, valid for a stationary complex or real process (Eq. 10-42 in [5]):

$$R_{\xi\xi}(\tau) \triangleq E[\xi(t+\tau)\xi^*(t)], \tag{9}$$

where $E[\cdot]$ denotes statistical average and $(\cdot)^*$ complex conjugation. We can decompose the signal $\xi$ in its quadrature components

$$\xi(t) = \xi_c(t) + i\xi_s(t). \tag{10}$$

As all involved processes are complex envelopes of pass-band processes assumed to be wide sense stationary (WSS), then the following correlation properties hold (Eq. 11-63 in [5])

$$\begin{aligned} R_{\xi_c\xi_c}(\tau) &= R_{\xi_s\xi_s}(\tau), \\ R_{\xi_c\xi_s}(\tau) &= -R_{\xi_s\xi_c}(\tau) \end{aligned} \tag{11}$$

and consequently, by applying Eqs. (9) and (10) and taking into account Eq. (11), we have

$$R_{\xi\xi}(\tau) = 2R_{\xi_c\xi_c}(\tau) - i2R_{\xi_c\xi_s}(\tau). \tag{12}$$

Furthermore, being $\xi_c, \xi_s$ real, the following holds

$$\begin{aligned} R_{\xi_c\xi_c}(-\tau) &= R_{\xi_c\xi_c}(\tau), \\ R_{\xi_c\xi_s}(-\tau) &= -R_{\xi_c\xi_s}(\tau). \end{aligned} \tag{13}$$

We will make use of the following normalised coefficients

$$\begin{aligned} \rho_\xi(\tau) &= \frac{2R_{\xi_c\xi_c}(\tau)}{P_\xi}, \rho_\xi(0) = 1, \\ \mu_\xi(\tau) &= \frac{2R_{\xi_c\xi_s}(\tau)}{P_\xi}, \mu_\xi(0) = 0, \end{aligned} \tag{14}$$

where $P_\xi$ is the power of the process, such that the autocorrelation of Eq. (9) is expressed by

$$R_{\xi\xi}(\tau) = P_\xi\left(\rho_\xi(\tau) - i\mu_\xi(\tau)\right). \tag{15}$$

We also define, for later use, the complex correlation coefficient

$$r_\xi(\tau) = |r_\xi(\tau)|e^{i\varphi_\xi(\tau)} = \rho_\xi(\tau) - i\mu_\xi(\tau) \tag{16}$$

with obviously

$$|r_\xi|^2 = \rho_\xi^2 + \mu_\xi^2 \leq 1. \tag{17}$$

## III. Mean of the Complex Ratio

Even though the expression of the mean of the complex ratio is available in literature [2], it is re-computed here to adapt it to the scenarios and terminology considered in this paper. Let's form the monopulse ratio based on Eqs. (2) and (3)

$$M(t) = \frac{\theta_s K_F e^{i\varphi_s}\Sigma(t) + n_\Delta(t)}{\Sigma(t) + n_\Sigma(t)}. \tag{18}$$

When applying the statistical expectation to the above, and after averaging with respect to $n_\Delta(t)$ (assumed to be statistically independent from $n_\Sigma(t)$), we immediately get

$$E[M(t)] = \theta_s K_F e^{i\varphi_s} E\left[\frac{\Sigma(t)}{\Sigma(t) + n_\Sigma(t)}\right]. \tag{19}$$

When comparing Eq. (19) with Eq. (6) and when taking into account Eq. (5), we have

$$\alpha = E\left[\frac{\Sigma(t)}{\Sigma(t) + n_\Sigma(t)}\right]. \tag{20}$$

From the above it is evident that, under the assumption of statistical independence between sum and difference noise, the reduction factor $\alpha$ applied to the noiseless complex monopulse ratio depends only upon the noise present in the sum channel. By taking into account Eq. (7), Eq. (20) becomes

$$\begin{aligned} \alpha = {} & E\left[\frac{\sqrt{P_c}e^{i\theta_c}}{\sqrt{P_c}e^{i\theta_c} + x + n_\Sigma}\right] \\ & + E\left[\frac{x}{\sqrt{P_c}e^{i\theta_c} + x + n_\Sigma}\right], \end{aligned} \tag{21}$$

having set $t = 0$ due to the wide sense stationarity of all involved processes, and with the understanding that $x, n_\Sigma$ are evaluated in the time origin. Let's tackle the first expectation in Eq. (21), and as a first step we average with respect to $\theta_c$.

$$\begin{aligned} & E_{\theta_c}\left[\frac{\sqrt{P_c}e^{i\theta_c}}{\sqrt{P_c}e^{i\theta_c} + x + n_\Sigma}\right] \\ & = \frac{1}{2\pi}\int_0^{2\pi} \frac{e^{i\theta_c}}{e^{i\theta_c} + \frac{x + n_\Sigma}{\sqrt{P_c}}} d\theta_c. \end{aligned} \tag{22}$$

By changing variable in the integral as follows: $e^{i\theta_c} = w$, and by applying the residue theorem it is immediate to verify that

$$E_{\theta_c}\left[\frac{\sqrt{P_c}e^{i\theta_c}}{\sqrt{P_c}e^{i\theta_c} + x + n_\Sigma}\right] = \begin{cases} 1 & |x + n_\Sigma| < \sqrt{P_c} \\ 0 & |x + n_\Sigma| \geq \sqrt{P_c} \end{cases}. \tag{23}$$

The remaining expectation with respect to $x, n_\Sigma$ immediately leads to

$$E_{x,n_\Sigma}\left[\begin{cases}1 & |x+n_\Sigma| < \sqrt{P_c} \\ 0 & |x+n_\Sigma| \geq \sqrt{P_c}\end{cases}\right] = 1 - e^{-\frac{P_c}{P_x+P_\Sigma}}. \qquad (24)$$

Let's now focus on the second expectation of Eq. (21). Again the average with respect to $\theta_c$ is easily conducted by use of residue theorem, leading to

$$E_{\theta_c}\left[\frac{x}{\sqrt{P_c}e^{i\theta_c} + x + n_\Sigma}\right] = \begin{cases}\dfrac{x}{x+n_\Sigma} & |x+n_\Sigma| > \sqrt{P_c} \\ 0 & |x+n_\Sigma| \leq \sqrt{P_c}\end{cases}. \qquad (25)$$

When performing the residual averaging with respect to $x, n_\Sigma$ we obtain the following

$$E_{x,n_\Sigma}\left[\begin{cases}\dfrac{x}{x+n_\Sigma} & |x+n_\Sigma| > \sqrt{P_c} \\ 0 & |x+n_\Sigma| \leq \sqrt{P_c}\end{cases}\right] = \frac{P_x}{P_x+P_\Sigma}e^{-\frac{P_c}{P_x+P_\Sigma}}. \qquad (26)$$

When using Eqs. (24) and (26) in Eq. (21) we finally obtain

$$\alpha = 1 - e^{-\frac{P_c}{P_x+P_\Sigma}}\left(\frac{P_\Sigma}{P_x+P_\Sigma}\right), \qquad (27)$$

which is in line with Eq. (72) of [2], which is more general in allowing a statistical dependence between sum and difference noise, and however considers a constant signal term. Fig. 3 visualises Eq. (27) as a function of the following two parameters

$$SNR = \frac{P_x+P_c}{P_\Sigma}, \qquad \lambda = \frac{P_x}{P_x+P_c}, \qquad (28)$$

which express the signal to noise ratio in the sum channel ($SNR$) and the relative proportion between Gaussian and monochromatic components in the signal part ($\lambda = 1$ corresponds to a purely Gaussian signal, $\lambda = 0$ to a residual carrier only). It is evident from Fig. 3 that, for a wide range of signal to noise ratios, the reduction is more pronounced for a pure Gaussian signal than for a pure monochromatic signal.

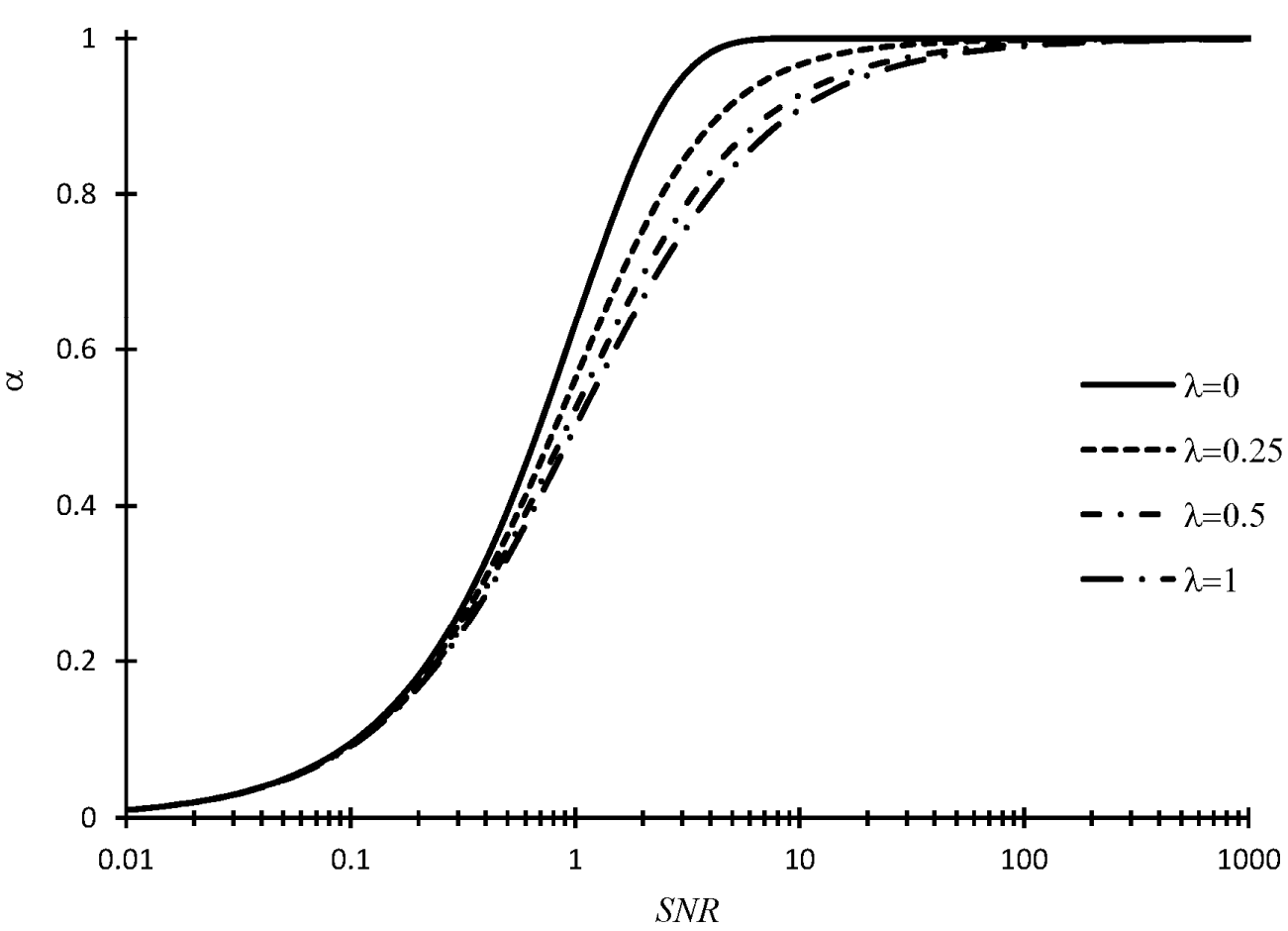

Fig. 3: reduction factor $\alpha$ as a function of the signal to noise ratio in the sum channel. The top curve corresponds to the case a pure monochromatic signal, whereas the bottom one is valid for a Gaussian signal. The curves in between address the simultaneous presence of both components.

## IV. Spectral density of the complex ratio

Let's now turn our attention to the spectral density of $\delta M(t)$, which can be obtained from its autocorrelation

$$R_{\delta M \delta M}(\tau) = E[\delta M(t+\tau)\delta M(t)^*]. \qquad (29)$$

Indeed the spectral density $S_{\delta M}(f)$ is obtained by Fourier transform of Eq. (29). As discussed in the introduction however, we are interested in the spectral density around the origin of the complex baseband, which means that we will compute

$$S_{\delta M}(0) = \int_{-\infty}^{\infty} R_{\delta M \delta M}(\tau) d\tau. \qquad (30)$$

As discussed in the introduction we are interested in computing the above quantities during tracking, when $\theta_s$ is vanishing (null tracking) and the mean of $M$ is zero, which, when taking into account Eq. (2) and (3), leads to

$$R_{\delta M \delta M}(\tau) = E\left[\frac{n_\Delta(t+\tau)}{\Sigma(t+\tau) + n_\Sigma(t+\tau)} \cdot \frac{n_\Delta(t)^*}{\Sigma(t)^* + n_\Sigma(t)^*}\right], \qquad (31)$$

which can be plugged in Eq. (30) to give, when using Eq. (7) and after setting $t = 0$ due to the wide sense stationarity of all involved processes

$$S_{\delta M}(0) = \int_{-\infty}^{\infty} E\left[\frac{n_\Delta(\tau)}{\sqrt{P_c}e^{i(2\pi f_c \tau + \theta_c)} + x(\tau) + n_\Sigma(\tau)} \cdot \frac{n_\Delta(0)^*}{\sqrt{P_c}e^{-i\theta_c} + x(0)^* + n_\Sigma(0)^*}\right] d\tau. \qquad (32)$$

We multiply and divide the above by the same exponential term $e^{-i2\pi f_c \tau}$ to get

$$S_{\delta M}(0) = \int_{-\infty}^{\infty} E\left[\frac{n'_\Delta(\tau)}{\sqrt{P_c}e^{i\theta_c} + x'(\tau) + n'_\Sigma(\tau)} \cdot \frac{n'_\Delta(0)^*}{\sqrt{P_c}e^{-i\theta_c} + x'(0)^* + n'_\Sigma(0)^*}\right] d\tau, \tag{33}$$

having defined the following signals.

$$\begin{aligned} n'_\Delta(t) &= n_\Delta(t)e^{-i2\pi f_c t}, \\ n'_\Sigma(t) &= n_\Sigma(t)e^{-i2\pi f_c t}, \\ x'(t) &= x(t)e^{-i2\pi f_c t}. \end{aligned} \tag{34}$$

The above corresponds to have frequency shifted all signal and noise spectra (residual carrier and Gaussian terms) by $-f_c$, thus positioning the residual carrier in the centre of the complex baseband. In the frame of the analysis we will make reference to the total Gaussian signal in the sum channel, including the signal component $x'(t)$ and the noise $n'_\Sigma(t)$

$$z(t) = z_c(t) + iz_s(t) = x'(t) + n'_\Sigma(t). \tag{35}$$

The process $z(t)$ is WSS, as it can be easily verified by computing $E[z(t_1)z(t_2)^*]$, and using Eqs. (34), (35) as well as the fact that $x(t), n_\Sigma(t)$ are assumed to be zero mean WSS processes. The autocorrelation of $z(t)$ can be expressed as

$$\begin{aligned} R_{zz}(\tau) &= 2R_{z_c z_c}(\tau) - i2R_{z_c z_s}(\tau) \\ &= P\big(\rho(\tau) - i\mu(\tau)\big), \end{aligned} \tag{36}$$

where

$$P = P_x + P_\Sigma \tag{37}$$

and with similar properties as presented in section II for the originating Gaussian signal. Furthermore we define also in this case the normalised correlation coefficient as follows

$$r(\tau) = |r(\tau)|e^{i\varphi(\tau)} = \rho(\tau) - i\mu(\tau). \tag{38}$$

Due to the assumed statistical independence between the noise on the difference channel and the signal and noise components in the sum channel, Eq. (33) becomes

$$S_{\delta M}(0) = \int_{-\infty}^{\infty} R_{n'_\Delta n'_\Delta}(\tau) R_{ss}(\tau) d\tau, \tag{39}$$

where we have denoted with $R_{ss}(\tau)$ the autocorrelation of the following complex signal

$$s(t) \triangleq \frac{1}{\sqrt{P_c} \cdot e^{i\theta_c} + z(t)}. \tag{40}$$

From now onwards we move our attention to the "difficult" term in Eq. (39), namely $R_{ss}(\tau)$, and we will re-discuss $S_{\delta M}(0)$ only at the end of the section. In order to simplify the terminology we define subscripts 1,2 for instances of the same variable separated by a time interval $\tau$, define $A = \sqrt{P_c}$, and re-write $R_{ss}(\tau)$ as follows

$$R_{ss}(\tau) = E\left[\frac{1}{A \cdot e^{i\theta_c} + z_{c,2} + iz_{s,2}} \cdot \frac{1}{A \cdot e^{-i\theta_c} + z_{c,1} - iz_{s,1}}\right]. \tag{41}$$

The computation of the above expectation requires the knowledge of the covariance matrix for the Gaussian column vector $\mathbf{z} = \{z_{c,1}, z_{s,1}, z_{c,2}, z_{s,2}\}'$ (the superscript $'$ denotes vector transposition), which, based on Eq. (36), is simply given by

$$\mathbf{R} = \frac{P}{2}\begin{bmatrix} 1 & 0 & \rho & -\mu \\ 0 & 1 & \mu & \rho \\ \rho & \mu & 1 & 0 \\ -\mu & \rho & 0 & 1 \end{bmatrix}. \tag{42}$$

The expectation (41) can be computed as follows

$$R_{ss}(\tau) = \frac{1}{2\pi}\int_0^{2\pi}\left[\int_{-\infty}^{\infty} \frac{p\big(z_{c,1}, z_{s,1}, z_{c,2}, z_{s,2}\big)}{A \cdot e^{i\theta_c} + z_{c,2} + iz_{s,2}} \cdot \frac{dz_{c,1}dz_{s,1}dz_{c,2}dz_{s,2}}{A \cdot e^{-i\theta_c} + z_{c,1} - iz_{s,1}}\right] d\theta_c, \tag{43}$$

where $p(\cdot)$ is the multivariate Gaussian probability density function associated to the vector $\mathbf{z}$. Before proceeding further let's apply the following unitary variable transformation to the inner integral

$$\mathbf{z} = \begin{bmatrix} \cos\theta_c & -\sin\theta_c & 0 & 0 \\ \sin\theta_c & \cos\theta_c & 0 & 0 \\ 0 & 0 & \cos\theta_c & -\sin\theta_c \\ 0 & 0 & \sin\theta_c & \cos\theta_c \end{bmatrix} \cdot \mathbf{w}, \tag{44}$$

such that the inner integral becomes, after simples computations

$$\begin{aligned} &\int_{-\infty}^{\infty} \frac{p\big(z_{c,1}, z_{s,1}, z_{c,2}, z_{s,2}\big)}{A \cdot e^{i\theta_c} + z_{c,2} + iz_{s,2}} \cdot \frac{dz_{c,1}dz_{s,1}dz_{c,2}dz_{s,2}}{A \cdot e^{-i\theta_c} + z_{c,1} - iz_{s,1}} \\ &= \int_{-\infty}^{\infty} \frac{p\big(w_{c,1}, w_{s,1}, w_{c,2}, w_{s,2}\big)}{A + w_{c,2} + iw_{s,2}} \cdot \frac{dw_{c,1}dw_{s,1}dw_{c,2}dw_{s,2}}{A + w_{c,1} - iw_{s,1}}, \end{aligned} \tag{45}$$

which is independent from $\theta_c$, and with $p(\cdot)$ being the same multivariate Gaussian probability density function introduced in Eq. (43) . Based on the above we can simply re-write Eq. (43) without the averaging with respect to $\theta_c$, and setting $\theta_c$ equal to zero.

$$R_{ss}(\tau) = \int_{-\infty}^{\infty} \frac{p(\mathbf{z})}{\left(A + z_{c,2} + iz_{s,2}\right)\left(A + z_{c,1} - iz_{s,1}\right)} d^4\mathbf{z}. \tag{46}$$

The multivariate density $p(\mathbf{z})$ is related to its characteristic function $c(\mathbf{x})$ by

$$p(\mathbf{z}) = \frac{1}{(2\pi)^4} \int_{-\infty}^{\infty} c(\mathbf{x}) e^{-i\mathbf{x}'\cdot\mathbf{z}} d^4\mathbf{x}, \tag{47}$$

where $\mathbf{x} = \{x_{c,1}, x_{s,1}, x_{c,2}, x_{s,2}\}'$, $\cdot$ denotes scalar product and where, for a zero mean Gaussian vector (Eq. (8-57) in [5])

$$c(\mathbf{x}) = e^{-\frac{1}{2}\mathbf{x}'\cdot\mathbf{R}\cdot\mathbf{x}}. \tag{48}$$

When using Eqs. (47) and (48) into Eq. (46) we have

$$\begin{aligned} R_{ss}(\tau) &= \frac{1}{(2\pi)^4} \int_{-\infty}^{\infty} e^{-\frac{1}{2}\mathbf{x}'\cdot\mathbf{R}\cdot\mathbf{x}} d^4\mathbf{x} \\ &\cdot \int_{-\infty}^{\infty} \frac{e^{-i\mathbf{x}'\cdot\mathbf{z}}}{\left(A + z_{c,2} + iz_{s,2}\right)\left(A + z_{c,1} - iz_{s,1}\right)} d^4\mathbf{z}. \end{aligned} \tag{49}$$

We focus now on the inner integral

$$I = \int_{-\infty}^{\infty} \frac{e^{-i\mathbf{x}'\cdot\mathbf{z}}}{\left(A + z_{c,2} + iz_{s,2}\right)\left(A + z_{c,1} - iz_{s,1}\right)} d^4\mathbf{z} \tag{50}$$

and perform the following change of variables

$$\mathbf{v} = A\mathbf{w} + \mathbf{z}, \tag{51}$$

where

$$\begin{aligned} \mathbf{v} &= \left\{v_{c,1}, v_{s,1}, v_{c,2}, v_{s,2}\right\}', \\ \mathbf{w} &= \{1,0,1,0\}', \end{aligned} \tag{52}$$

to get

$$I = e^{iA\mathbf{x}'\cdot\mathbf{w}} \int_{-\infty}^{\infty} \frac{e^{-i\mathbf{x}'\cdot\mathbf{v}}}{\left(v_{c,2} + iv_{s,2}\right)\left(v_{c,1} - iv_{s,1}\right)} d^4\mathbf{v}. \tag{53}$$

The integral in Eq. (53) is a 4-dimensional inverse Fourier transform which can be immediately computed, to give

$$I = -4\pi^2 \frac{e^{iA\mathbf{x}'\cdot\mathbf{w}}}{\left(x_{c,1} - ix_{s,1}\right)\left(x_{c,2} + ix_{s,2}\right)}. \tag{54}$$

The above result can also be obtained by changing to double polar coordinates in the integral of Eq. (53), by using Eq. (3.338) in [6] for the integration over the angle variables, and and by finally integrating over the radial variables. When plugging Eq. (54) into Eq. (49), after applying the following change of variables ($j = 1,2$)

$$\begin{aligned} x_{c,j} &= r_j \cos q_j, \\ x_{s,j} &= r_j \sin q_j, \end{aligned} \tag{55}$$

with $r_1, r_2 \in [0, \infty)$ and $q_1, q_2 \in [0, 2\pi)$, and by using Eq. (38) we obtain

$$\begin{aligned} R_{ss}(\tau) &= \frac{-4\pi^2}{(2\pi)^4} \int_0^{\infty} e^{-\frac{P}{4}\left(r_1^2 + r_2^2\right)} dr_1 dr_2 \\ &\cdot \int_0^{2\pi} e^{-\frac{|r|Pr_1r_2}{2}\cos(q_1 - q_2 + \varphi)} \\ &\cdot e^{iAr_1\cos q_1 + iAr_2\cos q_2} e^{i(q_1 - q_2)} dq_1 dq_2. \end{aligned} \tag{56}$$

We then use the following expansion

$$\exp(ia\cos x) = \sum_{n=-\infty}^{\infty} (i)^n J_n(a) \exp(inx), \tag{57}$$

where $J_v(\cdot)$ is the Bessel function of order $v$. By use of Eq. (57) we transform Eq. (56) into the following

$$\begin{aligned} R_{ss}(\tau) &= \frac{-4\pi^2}{(2\pi)^4} \int_0^{\infty} e^{-\frac{P}{4}\left(r_1^2 + r_2^2\right)} dr_1 dr_2 \\ &\cdot \sum_{n,m=-\infty}^{\infty} (i)^{n+m} J_n(Ar_1) J_m(Ar_2) \\ &\cdot \int_0^{2\pi} e^{-\frac{|r|Pr_1r_2}{2}\cos(q_1 - q_2 + \varphi)} \\ &\cdot e^{iq_1(n+1) - iq_2(m+1)} dq_1 dq_2. \end{aligned} \tag{58}$$

The inner integral of Eq. (58) can be computed easily, by lengthy computations around the integral representation of Eq. (8.431-5) in [6], to obtain

$$\begin{aligned} &\int_0^{2\pi} e^{-\frac{|r|Pr_1r_2}{2}\cos(q_1 - q_2 + \varphi)} \\ &\cdot e^{iq_1(n+1) - iq_2(m+1)} dq_1 dq_2 \\ &= 4\pi^2 (-1)^{n+1} I_{n+1}\left(\frac{|r|Pr_1r_2}{2}\right) e^{-i\varphi(n+1)} \delta_{nm}, \end{aligned} \tag{59}$$

where $\delta_{nm}$ is the Kronecker delta symbol and $I_v(\cdot)$ is the modified Bessel function of the first kind of order $v$. When plugging Eq. (59) into Eq. (58) we obtain

$$\begin{aligned} R_{ss}(\tau) &= \sum_{n=-\infty}^{\infty} e^{-i\varphi(n+1)} \int_0^{\infty} J_n(Ar_1) J_n(Ar_2) \\ &\cdot e^{-\frac{P}{4}\left(r_1^2 + r_2^2\right)} I_{n+1}\left(\frac{|r|Pr_1r_2}{2}\right) dr_1 dr_2. \end{aligned} \tag{60}$$

The integrals in Eq. (60) are solved by expanding the functions $I_v(\cdot)$ in power series and by using Eq. (6.631) in [6]

to resolve the resulting integrals, to finally get, after laborious manipulations

$$\begin{aligned} R_{ss}(\tau) &= \frac{1}{P_x+P_\Sigma}\sum_{n=0}^{\infty}\frac{(r(\tau)^*)^{n+1}\left(\frac{P_c}{P_x+P_\Sigma}\right)^n}{\Gamma(1+n)^2} \\ &\cdot \sum_{k=0}^{\infty}\frac{|r(\tau)|^{2k}\Gamma(n+1+k)}{\Gamma(k+1)(n+1+k)} \\ &\cdot\ {}_1F_1\left(1+k+n,1+n,-\frac{P_c}{P_x+P_\Sigma}\right)^2 \\ &+\frac{1}{P_x+P_\Sigma}\sum_{m=0}^{\infty}\frac{r(\tau)^m\left(\frac{P_c}{P_x+P_\Sigma}\right)^{m+1}}{\Gamma(2+m)^2} \\ &\cdot \sum_{k=0}^{\infty}\frac{|r(\tau)|^{2k}\Gamma(m+1+k)}{\Gamma(k+1)} \\ &\cdot\ {}_1F_1\left(1+k+m,2+m,-\frac{P_c}{P_x+P_\Sigma}\right)^2, \end{aligned} \tag{61}$$

where ${}_1F_1(\cdot,\cdot,\cdot)$ is the confluent hypergeometric function defined e.g. in Eq. (9.210-1) in [6], and where we have re-introduced $P_c = A^2$ and $P_x + P_\Sigma = P$. Eq. (61) constitutes the main output of this paper, and can be used in Eq. (39) to recover $S_{\delta M}(0)$.

We notice that if only Gaussian components are present in the sum channel ($P_c = 0$), the Eq. (61) reduces

$$[R_{ss}(\tau)]_{P_c=0} = -\frac{1}{R_{zz}(\tau)}\ln\left(1-\frac{|R_{zz}(\tau)|^2}{(P_x+P_\Sigma)^2}\right), \tag{62}$$

which is equivalent to Eq. (38) of [3], when assuming one target point and no offset. It is important to remark that both Eqs. (61) and (62) diverge for $\tau = 0$, which indicates that, irrespectively of the presence of a monochromatic signal of any finite amplitude, the process $s(t)$ of Eq. (40) has infinite power, which of course does not prevent $S_{\delta M}(0)$ of Eq. (39) from being finite.

## V. An application

Let's assume that the Gaussian random process on the sum channel (combination of signal and noise component) has a flat power spectrum over a bandwidth $W$ centred in the zero frequency of the complex baseband, with total power $P_x + P_\Sigma$. On the delta channel there is the same spectral structure with power $P_\Delta$. The monochromatic component (available only in the sum channel, during null tracking) is positioned at $f_c$ with power $P_c$. According to the procedure built in the previous section, we first shift all Gaussian and monochromatic spectra by $-f_c$ such that the autocorrelation of $z(t)$ and of $n'_\Delta$ of Eq. (36) and (39) are given by

$$\begin{aligned} R_{zz}(\tau) &= (P_x+P_\Sigma)\frac{\sin(\pi W\tau)}{\pi W\tau}e^{-i2\pi f_c\tau}, \\ R_{n'_\Delta n'_\Delta}(\tau) &= P_\Delta\frac{\sin(\pi W\tau)}{\pi W\tau}e^{-i2\pi f_c\tau}, \end{aligned} \tag{63}$$

which according to Eq. (36) implies that

$$\rho(\tau) - i\mu(\tau) = \frac{\sin(\pi W\tau)}{\pi W\tau}e^{-i2\pi f_c\tau}. \tag{64}$$

When inserting the above into Eq. (61) to obtain $R_{ss}(\tau)$, after plugging the results into Eq. (39) and when taking into account the expression of $R_{n'_\Delta n'_\Delta}(\tau)$ in Eq. (63) we obtain

$$S_{\delta M}(0) = \frac{P_\Delta}{W(P_x+P_\Sigma)}\chi\left(\frac{P_c}{P_x+P_\Sigma},\frac{f_c}{(W/2)}\right), \tag{65}$$

where we have defined the general function

$$\begin{aligned} \chi(a,b) &= \frac{2}{\pi}\sum_{n,k=0}^{\infty}\frac{a^n\Gamma(n+1+k)c_{2k+n+2}(nb)}{\Gamma(1+n)^2\Gamma(k+1)(n+1+k)} \\ &\cdot\ {}_1F_1(1+k+n,1+n,-a)^2 \\ &+\frac{2}{\pi}\sum_{m,k=0}^{\infty}\frac{a^{m+1}\Gamma(m+1+k)c_{2k+m+1}\big((m+1)b\big)}{\Gamma(2+m)^2\Gamma(k+1)} \\ &\cdot\ {}_1F_1(1+k+m,2+m,-a)^2 \end{aligned} \tag{66}$$

and where the coefficients $c$ have been defined as

$$c_n(m) \triangleq \frac{1}{2}\int_{-\infty}^{\infty}\left(\frac{\sin y}{y}\right)^n e^{imy}\,dy \tag{67}$$

and can be found according to Eq. (3.836-2) of [6]

$$\begin{aligned} c_n(m) &= c_n(-m) \\ &= \begin{cases} \dfrac{n\pi}{2^n}\displaystyle\sum_{v=0}^{\left\lfloor\frac{m+n}{2}\right\rfloor}\frac{(-1)^v(n+m-2v)^{n-1}}{v!\,(n-v)!} & 0\le m<n \\ 0 & m\ge n, n\ge 2 \\ \pi/4 & m=n=1 \end{cases}. \end{aligned} \tag{68}$$

It is to be noted that for the case of vanishing monochromatic signal one can use directly the closed form expression of $R_{ss}(\tau)$ of Eq. (62) to obtain the following limit case

$$\begin{aligned} \lim_{a\to 0}\chi(a,b) &= -\frac{1}{\pi}\int_{-\infty}^{\infty}\ln\left(1-\left(\frac{\sin(y)}{y}\right)^2\right)dy \\ &\cong 2.55. \end{aligned} \tag{69}$$

The function $\chi(a,b)$ was computed for various values of $b$ between 0 and 2 ($0 \le b \le 1$ corresponds to the realistic case of the monochromatic signal being within the noise spectrum), and for a set of values of $a$ between 0 and 3. After trials verifying proper numerical convergence of the sums in Eq. (66), the range from 0 to 20 was selected for the indexes $n, m$, whereas $k$ covered the range from 0 to 400. Furthermore the function $\chi(a,b)$ was also obtained as result of simulations where narrowband noise and signal components were digitally generated, the complex monopulse ratio formed according to Eq. (18), and its power spectrum estimated at the origin of the complex baseband. The results of theoretical computations (black points and dotted line) and simulations (red filled points) are presented in Fig. 4.

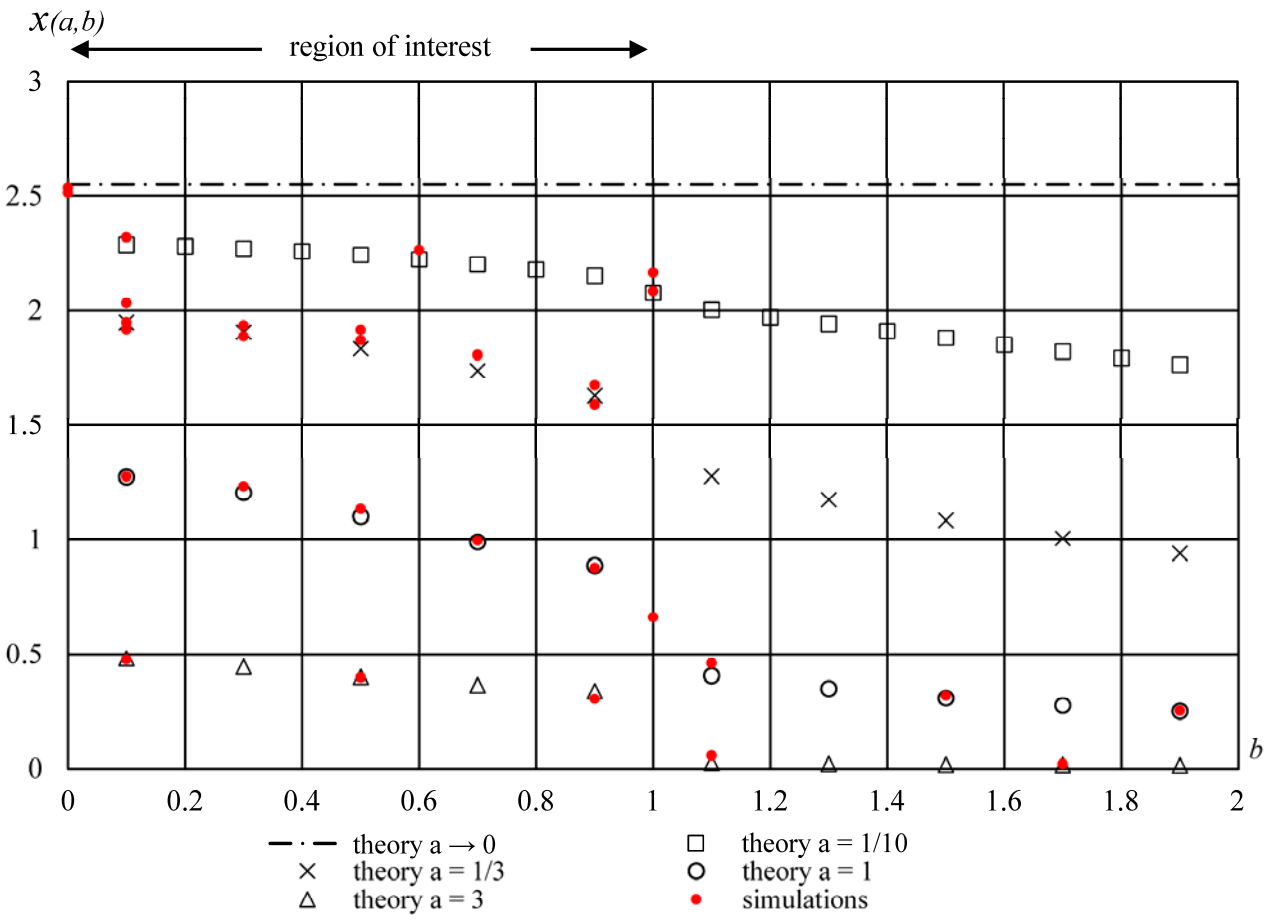

Fig. 4: computation of the normalized function $\chi(a,b)$ of Eq. (66) for various values of $a, b$. The red filled points represent simulations. It is to be noted that only values of $b$ up to 1 are of practical interest, as they correspond to a monochromatic signal within the noise spectrum. However the computations and simulations were extended to $b$ up to 2 to verify the agreement between theory and simulations.

## VI. Conclusions

The results of this paper can contribute to the design of closed loop tracking system for satellite missions, in those cases where the incoming signal cannot be modelled, when available at the input of the monopulse processor, as a simple Gaussian process. The Eq. (61), even though complex and involving double sums as well as evaluation of special function, can be practically used for computing the spectral density of the complex ratio fluctuations in the low frequency region. Such fluctuations, filtered by the time integration (4) and ultimately by the transfer function of the closed loop system, translate in a mechanical angular jitter of the antenna system during tracking and have a critical impact on the system performance.

## Acknowledgment

The author acknowledges the extremely valuable inputs from Dr. Yang Yang, of the National Laboratory of Radar Signal Processing, Xidian University, aiming at improving the content and formal correctness of the manuscript.

**Marco Lanucara** graduated from “La Sapienza” university of Rome in 1994 in electrical engineering. Since 2000 is a systems engineer at the European Space Agency, dealing with the design of satellite communications links for ESA near Earth and interplanetary missions. His main interest is about signal theory and its application to space communications and tracking.